\documentclass[twocolumn]{openjournal}
\usepackage{openjournal-submitted}
\usepackage{graphicx} 
\usepackage{multirow}

\usepackage[colorlinks,linkcolor=blue,citecolor=blue,urlcolor=blue ]{hyperref}
\usepackage[utf8]{inputenc}
\usepackage{float}
\usepackage{xcolor}
\usepackage{ulem}
\usepackage{amsmath}
\usepackage{xspace}
\usepackage{array,booktabs} 
\newcolumntype{P}[1]{>{\raggedright\arraybackslash}p{#1}} 

\newcommand{\enzoc}{\textsc{Enzo}}
\newcommand{\arepo}{\textsc{Arepo}}
\newcommand{\seedz}{\texttt{}{SEEDZ}}
\newcommand{\gwfish}{\texttt{}{GWFish }}

\newcommand{\msolar}{$\rm{M_{\odot}~}$}
\newcommand{\msolarc}{$\rm{M_{\odot}}$}

\newcommand{\rarepeak} {\textsc{Rarepeak}\xspace}
\newcommand{\NormalOne} {\textsc{Normal1}\xspace}
\newcommand{\NormalTwo} {\textsc{Normal2}\xspace}

\defcitealias{Prole2023}{LP23}
\defcitealias{Schauer2021}{AS21}

\begin{document}


\title{Hierarchical Black Hole Mergers at High Redshift: Predictions for LISA and LGWA from SEEDZ \vspace{-1cm}}
\author{Daxal H. Mehta$^{* \ 1}$}
\author{Lewis R. Prole$^{1}$}
\author{John A. Regan$^{1}$}
\author{R\"udiger Pakmor$^{2}$}
\author{Sophie Koudmani$^{3}$}
\author{Ricarda S. Beckmann$^{4}$}
\author{Michael Tremmel$^{5}$}
\author{John H. Wise$^{6}$}
\author{Martin A. Bourne$^{3, 10}$}
\author{Ralf S. Klessen$^{7,8}$}
\author{Simon C.~O. Glover$^{7,8}$}
\author{Martin G. Haehnelt$^{9, 10}$}
\author{John Brennan$^{1}$}
\author{Pelle van de Bor$^{1}$}
\author{Debora Sijacki$^{9,10}$}
\author{Paul C. Clark$^{11}$}
\email{$^*$email: daxal.mehta.2024@mumail.ie}
\affiliation{\\ \ \ \ \ $^{1}$ Centre for Astrophysics and Space Sciences Maynooth, Department of Physics, Maynooth University, Maynooth, Ireland.}
\affiliation{$^{2}$ Max-Planck-Institut f\"ur Astrophysik, Karl-Schwarzschild-Straße 1, D-85748, Garching, Germany}
\affiliation{$^{3}$ Centre for Astrophysics Research, Department of Physics, Astronomy and Mathematics, University of Hertfordshire, College Lane, Hatfield, AL10 9AB, UK}
\affiliation{$^{4}$ Institute for Astronomy, University of Edinburgh, Royal Observatory, Edinburgh EH9 3HJ, UK}
\affiliation {$^{5}$ School of Physics, University College Cork, College Road, Cork T12 K8AF, Ireland}
\affiliation {$^{6}$ Center for Relativistic Astrophysics, School of Physics, Georgia Institute of Technology, 837 State Street, Atlanta, GA 30332, USA}
\affiliation {$^{7}$ Universit\"{a}t Heidelberg, Zentrum f\"{u}r Astronomie, Institut f\"{u}r Theoretische Astrophysik, Albert-Ueberle-Stra{\ss}e 2, 69120 Heidelberg, Germany.}
\affiliation{$^{8}$ Universit\"{a}t Heidelberg, Interdisziplin\"{a}res Zentrum f\"{u}r Wissenschaftliches Rechnen, Im Neuenheimer Feld 225, 69120 Heidelberg, Germany} 
\affiliation{$^{9}$ Institute of Astronomy, University of Cambridge, Madingley Road, Cambridge CB3 0HA, UK}
\affiliation {$^{10}$ Kavli Institute of Cosmology, Cambridge, University of Cambridge, Madingley Road, Cambridge CB3 0HA, UK}
\affiliation {$^{11}$ Cardiff University School of Physics and Astronomy}


\begin{abstract}
  \noindent  We investigate the hierarchical assembly of black holes (BHs) in the early Universe using the next-generation SEEDZ cosmological zoom-in simulations. Our suite follows two independent regions (\rarepeak and \NormalOne) evolved with three different BH feedback prescriptions: no feedback (NoFB), weak feedback (WeakFB), and  standard thermal isotropic feedback (FidBH).  Focusing on BH mergers at $z \geq 10$, we explore merger-driven BH growth and detectability with next generation gravitational-wave observatories. We find that stronger AGN feedback models dramatically suppress hierarchical BH assembly compared to weaker feedback implementations. In the NoFB simulations, BHs frequently undergo repeated mergers, with some BHs experiencing more than 50 merger events within a few hundred Myr, gaining over $10^8 $ \msolar through mergers alone. In contrast, BHs in the FidBH simulations rarely undergo more than three mergers in this time. The NoFB implementation should be  seen as an upper limit to BH merger driven growth and beyond what is expected in the Universe. The suppression of BH mergers in the FidBH case is not driven by differences in galaxy assembly, BH occupation fractions, or indeed BH seed abundances, which remain broadly similar across the simulation suites. Instead, feedback disrupts the dense gas surrounding newly formed BH seeds, suppressing their early accretion-driven growth and preventing them from entering runaway hierarchical merger pathways. Our results demonstrate that early-Universe feedback subgrid prescriptions have a significant bearing on the hierarchical growth of BHs via both mergers and accretion, producing distinct high-redshift merger populations in our models. Coupled with seeding choices and other BH subgrid prescriptions this leaves us with multiple degeneracies that will need to be broken in advance of LISA and other next generation GW facilities.

\end{abstract}


 \maketitle
\section{Introduction} 
\label{sec:intro}
\noindent The discovery of supermassive black holes (BHs) (M$_{\rm BH} \gtrsim 10^6$ \msolarc) in the early Universe and therefore understanding their formation pathways has become one of the central challenges in modern astrophysics. Observations with the James Webb Space Telescope (JWST) are now revealing a population of active BHs at redshifts $z >6$, including systems with inferred black holes masses between $10^6-10^9$ \msolar within the first billion years of cosmic history \citep[e.g.][]{Maiolino2023, Kocevski2023, Matthee2024, Juodzbalis2025}. Explaining the rapid assembly of these objects remains difficult, particularly so for models where massive BHs originate from comparatively low-mass seeds and must grow efficiently through a combination of gas accretion and mergers \citep{Wise2008, Alvarez2009, Milosavljevic2009, Smith2018}.

The origin of these massive BHs remains an active area of research, with several competing formation pathways. The most widely studied scenario involves the remnants of Population III (PopIII) stars, which form from pristine gas in the first star-forming minihalos and leave behind BHs with masses ranging from a few solar masses up to a few hundred solar masses \citep[e.g.][]{Hirano2014, Clark2011b, Prole2022}. These BHs are commonly referred to as light seeds and are expected to be numerous throughout the early Universe \citep{Madau2001, Smith2018, Banik2019, Singh2023, Mehta2024, Regan2024a, Cammelli2025, Mehta2026}. Alternative pathways involve runaway stellar collisions in dense stellar clusters, where repeated mergers between massive stars and compact remnants can produce BHs with masses significantly larger than typical PopIII remnants \citep{Devecchi2009, Antonini2019, Rantala2025a, Rantala2025b,Paiella2026}. A third possibility is the ``direct collapse'' of massive gas clouds through an intermediate supermassive stellar or quasi-stellar phase\citep{Begelman2006,Wise2008, Regan2009,Fernandez2014, Latif2013b,Regan2014, Wise2019,Latif2022,prole2024a}. The pathways produce BHs with masses of $10^3-10^6$ \msolarc, often referred to as heavy seeds. 

Regardless of the initial seeding pathway, all proposed models face the challenge of growing BHs to the masses inferred by JWST observations within a few hundred million years of the Big Bang. Accretion is one method through which BHs can grow, however sustaining Eddington-limited accretion rate over extended periods is difficult, particularly for low-mass seeds that are susceptible to stellar and radiative feedback \citep{Wise2008, Alvarez2009, Milosavljevic2009, Smith2018}. Consequently, many models invoke episodes of super-Eddington accretion to reproduce the observed massive BH populations \citep{Lupi2016, Shi2023, Prole2026b}. While such growth may occur under favorable conditions, its frequency and duration remains uncertain \citep{Mehta2024, Mehta2026}. Hierarchical BH mergers provide an alternative and complementary growth channel that is not subject to the Eddington limit nor to the negative effects of accretion feedback. Repeated mergers can rapidly increase BH masses while simultaneously producing a population of gravitational wave (GW) sources that may be detectable by future observatories \citep{Amaro-Seoane2023, Puntaro2014, Harms2021}.

Massive BH mergers are expected to play an important role in the assembly of massive BHs in the high-redshift Universe \citep{Bhowmick2024a, Bhowmick2025a, Bhowmick2026}. In hierarchical structure formation models, galaxies grow through repeated mergers and accretion events, naturally driving their central BHs toward galactic centers. If these black holes efficiently sink and coalesce, repeated BH mergers may offer a pathway to increasingly massive BHs at early cosmic times without the deleterious effects of accretion feedback.

Observationally, upcoming missions and detectors  such as the Laser Interferometer Space Antenna (LISA: \citealt{Amaro-Seoane2023}), the Einstein Telescope (ET: \citealt{Puntaro2014})\footnote{The mergers considered here lie outside the sensitive frequency range of the ET, and hence we do not present an ET detectability analysis.}, and the Lunar Gravitational Wave Observatory (LGWA: \citealt{Harms2021}) will open a new observational window on this epoch, by directly probing massive BH mergers across cosmic history. The properties of these high-redshift GW populations should therefore provide valuable constraints on early BH formation and growth pathways. Realizing this potential, however, will require high-fidelity astrophysical models: only by combining GW observations and statistics with theoretical predictions can an accurate picture of the massive black hole population and its environment be obtained \citep[e.g.][]{Hartwig2018}.

 Despite their importance, the efficiency of hierarchical BH mergers in the early Universe remains poorly understood and poorly modeled. BH growth is strongly regulated by stellar and BH feedback, both of which can alter the dense gaseous environments surrounding young BH seeds \citep[e.g.][]{Dubois2015,Habouzit2016,Habouzit2017,Regan2019, Prole2026b}. Feedback may suppress gas accretion, disrupt nuclear gas reservoirs, and modify the ability of BHs to efficiently grow and participate in repeated merger events. At the same time, feedback also regulates galaxy evolution itself, making it difficult to disentangle the direct and indirect effects on BH assembly. As a result, understanding how feedback shapes the merger-driven growth of BH populations requires self-consistent cosmological simulations capable of simultaneously modeling star formation, BH seeding, accretion, feedback, and galaxy assembly. 

 In this work, we investigate the hierarchical assembly of BHs using the next generation \seedz\ cosmological zoom-in simulation suite. Our simulations follow two independent environments, including one overdense rare-peak region and one mean density regions, each evolved with three different black hole feedback prescriptions: no feedback (NoFB), weak feedback (WeakFB) and fiducial feedback (FidBH; where NoFB signifies no massive BH accretion feedback, WeakFB signifies a weakened massive BH accretion feedback prescription at high Eddington rates and FidBH stands for our canonical thermal feedback prescription -- see \S \ref{Sec:MBHFeedback} for details). In this study we only analyze the \rarepeak and \NormalOne regions (six simulations in total). \\
 \indent The \seedz\ model \citep{Prole2026b, Prole2026a, Prole2026c, Mehta2026} incorporates physically motivated prescriptions for both light and heavy BH seed formation, allowing us to study the interplay between seeding, accretion, and mergers within a unified framework. The goal of the simulation suite is primarily to understand massive BH formation and growth at high redshift. 

 This paper is organized as follows. In \S \ref{sec:method}, we describe the numerical methodology, simulation setup, BH seeding and feedback prescriptions employed in \seedz. \S \ref{sec:box}
describes our initial condition and the simulation suite. In \S \ref{sec:results}, we present the merger statistics, hierarchical merger trees, and growth histories of BHs across the simulation suites. In \S \ref{Sec:GWDetections} we analyse the expected number of detections by LISA and the LGWA from this epoch. In \S \ref{sec:discussion}, we discuss the implication of our results for early massive BH assembly and future gravitational wave observations with LISA. Finally, we present our conclusions in \S \ref{sec:conclusions}.

\section{Numerical method}
\label{sec:method}
The simulation framework employed for this study utilizes the same methodology as the original \seedz\  simulations \citep{Prole2026a, Prole2026b}. We therefore avoid repeating a detailed description here and instead describe only the methodology pertinent to this study as well as the datasets used here.

\subsection{Seeding Black Holes and Stars} 
\label{Sec:seeding}

\noindent Stars and BHs are modeled as sink particles, building on the scheme introduced in Arepo by \citet{Wollenberg2020} and \citet{Tress2020}. The formation of these particles is triggered in high density gas when a gas cell has reached the minimum size ($\sim$ 1 pc physical) and can no longer refine further. These particles model the behavior of physical processes that occur below the resolution limit. In our \seedz\ model sink particles can form when the following criteria are met:

\begin{itemize}\setlength\itemsep{0.2em} 
    \item The cell density exceeds the local Jeans density defined by the Truelove criterion,
\begin{equation}
    \rho > \rho_{\rm J} = J^2 \frac{\pi c_s^2}{G\Delta x^2},
            \label{eq:Jeans_density}
\end{equation}
 where $\Delta x$ is the local cell length, $c_s$ is the local sound speed and $J=0.25$ is a constant \citep[see][]{Krumholz2004}. 
    \item The gas within the accretion region (a sphere of radius 5 times the minimum cell length), around the gas cell is gravitationally bound and converging.
    \item The host cell is not within the accretion region of another sink particle
\end{itemize}

Once the above conditions are satisfied, the gas cell is treated as a candidate sink particle and we check additional conditions to determine the type of sink particle that is formed. Our strategy here is based on a scheme previously employed in \enzoc\ \citep{Enzo-Collaboration2014, Brummel-Smith2019, Regan2018, Regan2018a, Regan2020}. We identify three distinct sink particle types, namely: population III stars (PopIII), population II star clusters (PopII), and massive black holes (BHs).  We determine the candidate sink particle's identity using the following criteria in order:

\begin{itemize}
  \item If the mass flux, calculated by summing the flux from all cells within the accretion region, onto the candidate sink particle is above a threshold of 1 \msolar yr$^{-1}$, a massive BH is formed \citep{Regan2023a}. The initial mass
  of this massive BH is sampled from a power-law initial mass function (IMF) with lower and upper mass limits of
  $5\times10^3$~\msolar\ and $10^{5}$~\msolar\ and
  a slope of -0.5. If the host gas cell doesn't have sufficient mass to create the required massive BH, then we take 90\% of the host gas cell's mass and set that as the massive BH's initial mass. 

    \item If the mass flux is below the massive BH threshold and the metallicity is greater than $10^{-4} \ \rm{Z}_{\odot}$, a PopII star cluster particle is formed. We populate the  particle with stars sampled from a Kroupa IMF \citep{kroupa2001, Clark2008, Dopcke2011, Dopcke2013}. Assuming a star formation efficiency of $\epsilon = 0.1$, we form a stellar particle of mass $\epsilon \rm{M_{gas\_cell}}$.

    \item  If both of the previous checks are unsuccessful, we check whether the $\rm{H}_2$ fraction is greater than $5\times 10^{-4}$. If it is, a single PopIII star is formed. The star's mass is sampled from a top heavy IMF,
\begin{equation}
    \rm{\frac{dN}{d \log{M}} \propto  M^{-1.3} exp \left[ -\left( \frac{M_{char}}{M} \right)^{1.6} \right] dM},
\end{equation}
where $\rm{M_{char}}$ is a characteristic mass of 20 \msolarc. The IMF has a lower and upper bound of 1 and 300 \msolarc, respectively. If the sampled stellar mass is higher than the mass of the host gas cell, we do not form a PopIII particle. See also \citealt{Prole2026b, Prole2026a} for additional details. 
\end{itemize}
In cases where none of the above checks are successful then no sink is formed. 

\subsection{Supernova Feedback}
\label{sec:SNe}
\noindent Each PopIII star is assigned a stellar lifetime according to its mass \citep{Maeder2009}. PopII star cluster particles are governed by an IMF and the stars with masses in excess of 8 \msolar are assigned a stellar lifetime and depending on their mass (see the following paragraphs) may experience a supernova event similar to PopIII stars. Once the star's lifetime (PopII and PopIII) is exceeded, its fate is decided as follows:

   \begin{itemize}
      \item Stars with mass between 11 and 20 \msolar  undergo a core collapse SN, ejecting $E_{\rm SN} = 10^{51} $ erg of energy in to the surrounding environment. These stars leave behind a helium core whose mass is calculated as  \citep{Nomoto2006}
      \begin{equation}
    m_{\rm He} = 0.1077 +  0.3383 (m_{\star}-11) \: {\rm M}_{\odot},
    \label{eq:He_core_small}
    \end{equation} 
    where $m_{\star}$ is the stellar mass in units of M$_\odot$.  
    
    \item Stars between 20 and 40 \msolar undergo a hypernova, whose explosion energy and helium core masses are interpolated from a fitting formula \citep{Nomoto2006}. 
    
    \item In the mass range 40-140 \msolarc, the stars experience a direct collapse into a BH and exhibit no SN explosion. 

    \item Stars with masses between 140-260 \msolar experience a pair-instability SN. We calculate their helium core mass and supernova energy as
      \begin{equation}
      m_{\rm He} = \frac{13}{24} (m_{\star} - 20)
      \label{eq:He_core_large}
      \end{equation}
      \begin{equation}
      E_{\rm SN} = \left[5 + 1.304 (m_{\rm He} - 64) \right] \times 10^{51} \: {\rm erg}.
      \label{eq:SN_E51}
      \end{equation}
    These particles do not leave behind a remnant and are completely disrupted by the SN.
    
    \item Finally, stars more massive than 260 \msolar also experience direct collapse into a BH.
    \end{itemize}

For  the PopIII stars, the remnant helium core becomes a BH while we do not form any BHs from the PopII stellar clusters. Allowing the BHs formed within PopII star clusters to evolve and potentially grow with time is an area for future improvement within the SEEDZ model. For both PopII and PopIII stars, we also return the mass of the ejected shell from each SNe event to the surrounding gas. At the same time, we inject metals of the shell mass to the gas from both stellar types.

The initial (Calibration) \seedz~ simulations \citep{Prole2026b, Prole2026a}  employed a momentum-driven SN feedback modeled designed to account for the unresolved momentum injection associated with the late stages of SN remnant evolution. During the development of the SEEDZ framework and following these initial calibration simulations, this prescription was revisited and refined following indications that the adopted implementation injected excessive energy into the surrounding gas when we ionized the entire injection region to $10^4$ K. Thus, we adopted a thermal energy injection model for the simulations presented in this work. 

The thermal energy, following each SNe explosion, is injected in a region of ten times the  minimum cell diameter ($\sim 20$ pc). The energy is divided into the cells in these region with a mass-weighing scheme \citep[see][]{Prole2026b, Prole2026a}. 

\begin{table*}
    \centering
    \begin{tabular}{|c|c|c|c|c|c|c|}\toprule
         Simulation &  Redshift & Feedback & Total BHs & BHs with Mergers & Total Mergers & Max BH Mass [\msolarc]\\\midrule
         RP-NoFB   & 15.00 & None     & 10428 & 468 & 1980 & $1.59 \times 10^8$\\
         RP-WeakFB & 15.00 & Weak     & 10082 & 204 & 558  & $2.24 \times 10^7$\\
         RP-FidFB  & 15.52 & Fiducial & 10000 & 23  & 29   & $1.27 \times 10^6$\\
         N1-NoFB   & 10.00 & None     & 3699  & 304 & 2573 & $2.60 \times 10^8$\\
         N1-WeakFB & 10.27 & Weak     & 4151  & 221 & 504  & $1.43 \times 10^7$\\
         N1-FidFB  & 10.00 & Fiducial & 6819  & 22  & 51  & $2.60 \times 10^6$\\
        \bottomrule\end{tabular}
    \caption{Summary of Simulations. C1: Names of the simulation, C2: The final redshift. C3: Feedback implementation. C4: Total number of BH present at the end of the simulation. C5: Total number of BHs that underwent merger event. C6: Total number of BH merger events. C7: Mass of the largest BH formed in the simulation.}
    \label{tab:Simulations}
\end{table*}

\subsection{Black Hole Accretion}
\noindent  Once formed, BHs grow through gas accretion and mergers. Gas accretion is computed within a spherical accretion region extending to five times the minimum cell size ($\sim 5$ physical pc radius). We employ the Bondi-Hoyle-Lyttleton formalism \citep{Bondi1952}, following the implementation described in \citet{Krumholz2004, Krumholz2006} and adopted in previous \seedz\ studies. The local gas density, sound speed, and relative velocity are estimated from gas cells within the accretion region using a distance-weighted kernel. This approach allows the accretion model to smoothly transition between regimes where the Bondi radius is either resolved or unresolved by the numerical grid.

To account for the turbulent nature of the interstellar medium, the accretion rate is further modified using the local gas vorticity following \citet{Krumholz2006}. Gas with significant rotational support is less efficiently accreted, reducing the growth rate relative to the classical Bondi estimate. The resulting accretion rate is used to remove gas from the surrounding gas cells while conserving linear momentum between gas and the BH particle. At the same time, the BH position is updated according to the center-of-mass motion of the system.,

\subsection{Black Hole Accretion Feedback} \label{Sec:MBHFeedback}
\noindent Accreting BHs deposit thermal energy into the surrounding gas, providing a self-regulating feedback mechanism that can influence both BH growth and the evolution of the host galaxy. In the \seedz~ model we inject thermal energy isotropically into gas cells within the accretion region according to
\begin{equation}
    E_{\rm FB} = \epsilon f_c  M_{\rm acc}c^2,
\end{equation}
where $M_{\rm acc}$ is the mass accreted during the timestep, $f_c = 0.05$ \citep{Booth2009} is the thermal coupling factor, and $\epsilon$ is the radiative efficiency. For sub-Eddington accretion, we calculate the radiative efficiency for a moderately spinning BH (a=0.7), which keeps $\epsilon \sim 0.1$. At super-Eddington accretion rates, the radiative efficiency is reduced following the slim-disk prescription of \citet{Madau2014}, which accounts for photon trapping and the saturation of the emergent luminosity ($\rm{L_{SE}}$).  We then calculate the radiative efficiency from $\rm{L_{SE}}$, 
\begin{equation}
\epsilon  = L_{\rm SE} / (\dot{M}_{\rm BH} c^2).
\end{equation}
\subsubsection{Weak Feedback Model}

\noindent In addition to the fiducial (FidBH) feedback model described above, we also consider a reduced-feedback variant (WeakFB) designed to bracket the uncertainty associated with BH feedback in dense, gas-rich environments where feedback may be contained due to photon trapping or other environmental effects. In this model, thermal feedback is only injected during sub-Eddington accretion episodes and we therefore assume feedback-free accretion in the super and hyper-Eddington regimes. We therefore assume that super-Eddington accretion flows are radiatively inefficient, with a large fraction of the liberated energy being advected into the BH rather than escaping into the surrounding medium. As a result, no thermal feedback is injected when the accretion rate exceeds the Eddington limit. 

Furthermore, in this scheme feedback energy is not deposited into low-density gas cells ($n < 1 {\rm cm}^{-3}$). This modification prevents thermal energy from being injected into regions where it would have little dynamical impact, thereby restricting feedback to dense gas that is more directly coupled to the BH growth process. The WeakFB model therefore provides an intermediate case between no feedback (NoFB) simulations and FidBH simulations, allowing us to investigate the role of different BH feedback prescriptions in regulating accretion and hierarchical BH assembly. 

\subsection{Dynamical Friction}

\noindent Massive black holes moving through a background distribution of dark matter and stars experience dynamical friction, a drag force arising from the gravitational wave generated by their motion \citep{Chandrasekhar1943, Binney2011}. This process removes orbital energy and angular momentum, causing massive objects to migrate toward the centers of their host gravitational potentials. In cosmological simulations, however, the finite mass resolution of dark matter particles limits the accurate treatment of dynamical friction. The large particle masses can lead to artificial two-body heating, which is typically mitigated by introducing a gravitational softening length below which gravitational forces are suppressed \citep{Governato1994, Kazantzidis2005}. While this reduces numerical noise, it also weakens the unresolved contribution of dynamical friction on scales smaller than the softening length. To compensate for this missing force, we apply an additional dynamical friction correction to black hole particles, accounting for interactions that occur within the unresolved volume
\begin{equation}
V_{\rm s} = \frac{4}{3}\pi r_{\rm s}^{3},
\end{equation}
where $r_{\rm s}$ is the gravitational softening length.

Our implementation follows the approach of \citet{Tremmel2015}, adapted for the \textsc{arepo2} framework. A particular challenge in \arepo\ arises from its hierarchical timestepping scheme, in which only a subset of particles are active on a given timestep. To obtain a consistent estimate of the local dark matter distribution surrounding each black hole, nearby dark matter particles are identified during global synchronization steps and assigned to the same timestep bin as the black hole. This ensures that the local environment used in the dynamical friction calculation evolves consistently with the black hole itself.

We assume that the dark matter velocity distribution within the softening radius is isotropic and that the classical Chandrasekhar formalism applies. Under these assumptions, only background particles moving more slowly than the black hole contribute to the drag force, yielding an acceleration
\begin{equation} 
a_{\rm DF} = - 4 \pi G^2 M_{\rm BH} \rho(<v_{\rm BH})\ln{\Lambda} \frac{v_{\rm BH}}{|v_{\rm BH}^3|} 
\end{equation}
where $\rho(<v_{\rm BH})$ is the density of dark matter particles with speeds lower than that of the black hole, estimated from particles contained within the smoothing volume, and $v_{\rm BH}$ is the black hole velocity. The Coulomb logarithm is defined as $\Lambda = \frac{b_{\rm max}}{b_{\rm min}}$, where $b_{\rm max}$ and $b_{\rm min}$ are the maximum and minimum impact parameters, respectively. Since gravitational interactions are already resolved beyond the softening scale, we adopt $b_{\rm max} = r_{\rm s}$ thereby restricting the correction to unresolved scales. The minimum impact parameter is taken to be the black hole's gravitational influence radius,
\begin{equation} 
b_{\rm min} = \frac{GM_{\rm BH}}{v_{\rm BH}^2}. 
\end{equation}

In the present work, the correction is computed using only the dark matter component. Dynamical friction arising from stars will be added in the next iteration of the \seedz~ model. The correction is applied only to black holes with masses exceeding five times the dark matter particle mass. Below this threshold, the black hole dynamics become increasingly sensitive to stochastic interactions with individual dark matter particles, causing the Chandrasekhar estimate to become noisy and unreliable \citep{Tremmel2015}. As the black hole grows, its influence radius increases and may eventually exceed the adopted maximum impact parameter. In this regime, the relevant scales are already numerically resolved and the additional correction is switched off.

Accurately capturing dynamical friction is particularly important for studies of black hole mergers, as it governs the orbital decay of black holes following galaxy mergers and influences their ability to participate in repeated hierarchical assembly.

\subsection{Stellar and Black Hole Mergers}
\label{sec:mergers}
\noindent Since the spatial scales associated with the final stages of BH binary evolution are unresolved in cosmological simulations, mergers are implemented through a sub-grid model that identifies pairs of particles likely to coalesce. Two BH particles are allowed to merge if all of the following conditions are satisfied:

   \begin{itemize}
      \item They lie within each other’s accretion radius ($\lesssim 5$ pc (physical)). 
      \item The BHs are moving towards each other.
      \item Their relative acceleration is negative.
      \item The pair is gravitationally bound.
   \end{itemize}

These criteria are designed to ensure that only particles undergoing genuine orbital decay are merged, while minimizing spurious mergers arising from chance encounters. Since BH particles do not carry thermal information, the gravitational binding criterion is evaluated by comparing the gravitational potential energy of the pair to its kinetic energy.

When a merger occurs, the more massive particle is retained and inherits the mass and linear momentum of the secondary particle. The position of the surviving particle is updated to the center of mass of the system, ensuring conservation of mass and linear momentum. BH binary mergers are assumed to coalesce instantaneously once the merger criteria are satisfied. Delays associated with unresolved binary hardening and gravitational-wave inspiral are not explicitly modeled in \seedz.

\section{Initial Conditions}  
\label{sec:box} 
The simulations presented in this work are based on the next-generation \seedz\ cosmological zoom-in simulation suite. The parent simulation box is the same as in \citet{Prole2026a}, initialized at $z=127$ with a side length of 40 cMpc h$^{-1}$, created through MUSIC \citep{Hahn2011} with cosmological parameters $\Omega_0=0.315$, $\Omega_\Lambda=0.685$, $\Omega_B=0.0486$ and $h=0.674$ \citep{planck-collaboration2020}.  The parent simulation contains dark matter only and is evolved to $z=10$ in order to identify candidates regions for high-resolution re-simulation.

In the calibration runs of the \seedz\ simulations, we zoomed in on one overdense rare-peak region (\rarepeak) and two approximately mean-density regions (\NormalOne and \NormalTwo). The \rarepeak region was centered on the most massive halo within the parent simulation at $z=10$, while the \NormalOne and \NormalTwo regions are selected to contain halos with masses approximately $10^{10}$ \msolar at the same redshift, representative of typical environments at high redshifts. In the calibrations runs, the box length on a side was 6 cMpc h$^{-1}$, with a total volume of 216 (cMpc)$^3$ h$^{-3}$.

In this next-generation of \seedz\ simulations, we use a smaller region of space (due to higher resolution and computational constraints) but keep the zoom-in centered on the same halos as the calibrated runs. To construct the zoom-in initial conditions, particles within a radius of 0.25 physical Mpc for the \rarepeak region and 0.5 physical Mpc (at $z=10$) for the \NormalOne and \NormalTwo regions are traced back from $z=10$ to their corresponding Lagrangian volumes at $z=127$. These traced regions are then regenerated at higher resolution using MUSIC within additional refinement levels applied to the enclosed ellipsoidal Lagrangian regions. The resulting refined regions have volumes of $1.81 \times 1.68\times 1.55$ (cMpc)$^3$ h$^{-3}$ for \rarepeak, $1.25 \times 2.02 \times 1.21$ (cMpc)$^3$ h$^{-3}$ for \NormalOne and $1.43 \times 1.54 \times 1.14$ (cMpc)$^3$ h$^{-3}$ for \NormalTwo. As previously discussed the \NormalTwo simulations have not yet reached z = 10 and are not part of this analysis. 

The final zoom simulations achieve a dark matter particle mass resolution of $10176.9 \ \rm{h}^{-1}$ \msolarc. The minimum gas cell radius reached 1 physical pc, while the minimum gravitational softening lengths for dark matter and gas are 47 physical pc and 3 physical pc, respectively. 

Each zoom-in region is evolved using three different BH feedback prescriptions: NoFB, WeakFB, and FidBH, resulting in a total of six simulations. Their information can be found in Table \ref{tab:Simulations}. The identical initial conditions employed across the feedback variants allow us to isolate the impact of BH feedback on the hierarchical assembly and merger histories of BHs.

\section{Results}

\label{sec:results}

\noindent In this work, we investigate the role of BH mergers in the assembly of the BH populations during the first few hundred million years of cosmic history. Using the next-generation SEEDZ simulation suite, we quantify the contribution of mergers to BH growth, examine how different feedback prescriptions regulate hierarchical BH assembly, and explore the observational consequences for future gravitational wave missions such as LISA and LGWA. A particular focus of this study is determining whether mergers provide an important growth channel for the massive BHs in the early Universe and how feedback modifies this pathway.

We analyze a suite of six cosmological zoom-in simulations consisting of two independent environments (\rarepeak and \NormalOne) evolved with three BH feedback prescriptions (NoFB, WeakFB, and FidBH). A summary of the simulations is provided in Table \ref{tab:Simulations}. The \rarepeak simulations were evolved to $z\sim15$, while the \NormalOne simulations reach lower redshifts of $z\sim10$.  Despite terminating at earlier cosmic times, the overdense \rarepeak region hosts the largest BH population, containing approximately $\sim 10000$ BHs, compared to $\sim4\times10^3-7\times10^3$ BHs in the mean-density (\NormalOne) region.

A strong dependence on feedback strength emerges when investigating basic summary statistics from each simulation. In the NoFB simulations (across both the \rarepeak and \NormalOne volumes), between 300 and 470 BHs (from $6.25$ \msolar to $> 10^8$ \msolarc) undergo at least one merger event, whereas the corresponding number decreases to only 200-220 and 22-23 in the WeakFB and FidBH runs, respectively. This reduction occurs despite the total number of BHs remaining less than a factor of 2 across the different feedback models. The impact on massive BH growth is even more dramatic; the most massive massive BHs in the NoFB simulations reach masses of $\sim  1.6-2.6\times 10^8$ \msolarc, while the most massive BHs in the WeakFB and FidBH simulations typically remain below a few times $10^7$ \msolar and a few times $10^6$ \msolar - an order of magnitude difference. These trends suggest that feedback does not merely regulate BH accretion, but fundamentally alters the merger-driven assembly pathway for growing massive BHs. In the following sections, we investigate the origin of these differences and quantify how feedback reshapes the hierarchical growth of massive BHs via mergers.
\begin{figure}
    \centering
    \includegraphics[width=\linewidth]{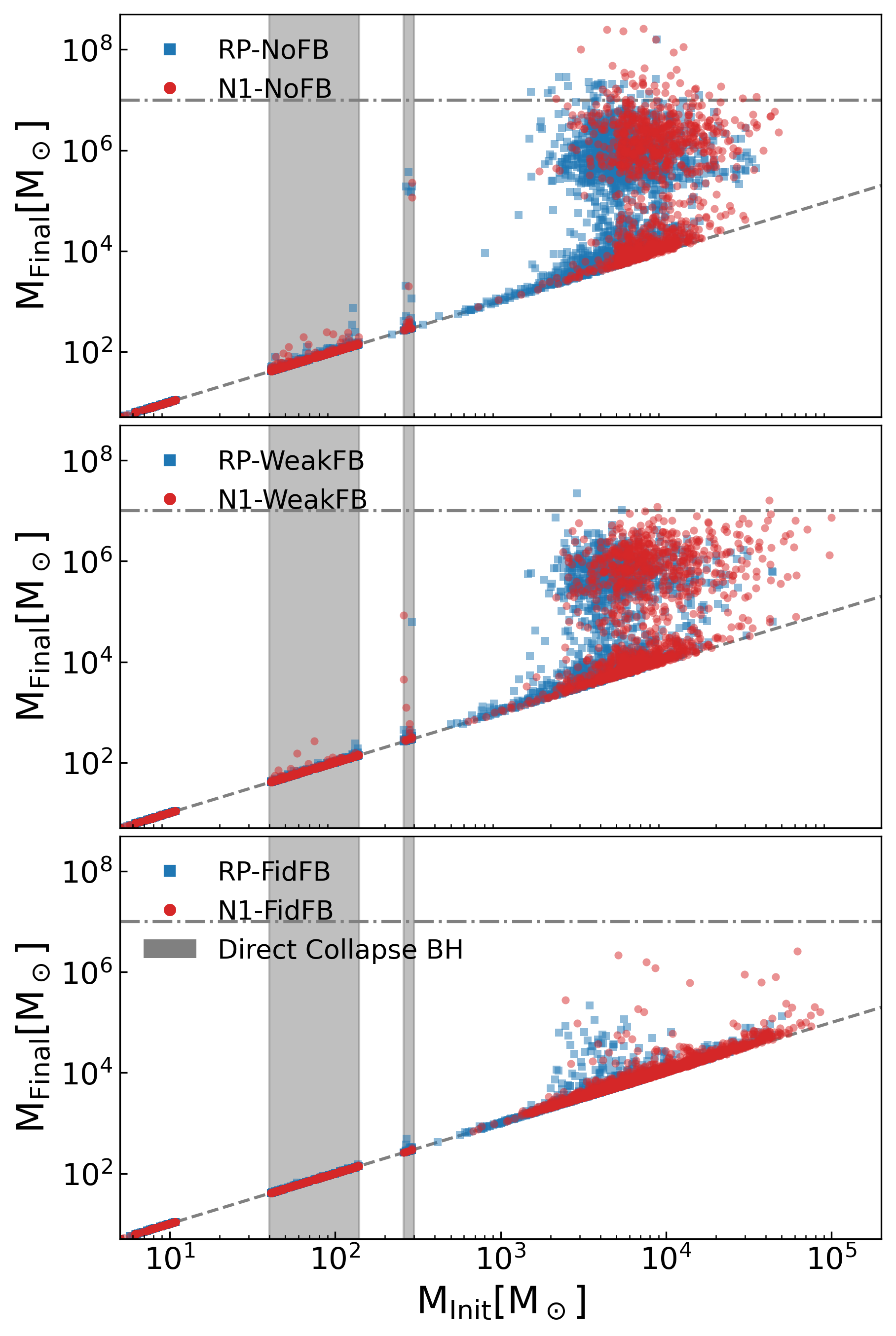}
    \caption{Final masses of BHs as a function of the initial mass for all six simulations. The \rarepeak region is showed in blue and \NormalOne region is showed in red. The simulation are divided in three panels according to feedback prescriptions. We also show a gray band where light seed black holes form through direct collapse.}
    \label{fig:Mi_Mf}
\end{figure}
\begin{figure*}
    \centering
    \includegraphics[width=0.8\linewidth]{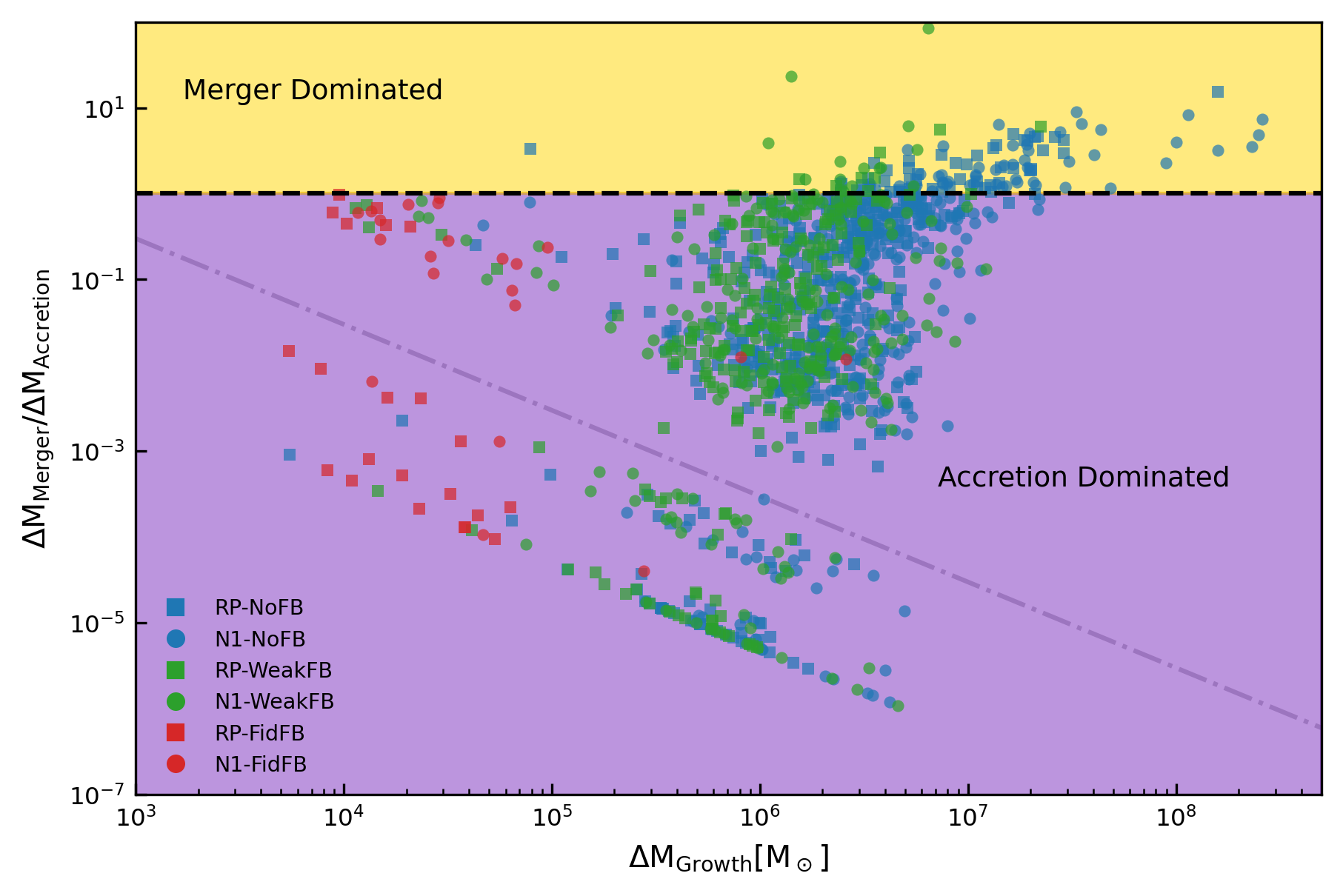}
    \caption{Fractional mass gained through mergers and accretion as a function of the final mass of the BHs. The different region are shown by different marker styles and feedback prescriptions have different colors. The horizontal dashed line marks equal growth from both channels and dotted-dashed line shows BHs that underwent only light seed mergers. Black holes sitting below the dotted-dashed line only experienced mergers with light seed BHs. Overall we find that the majority of BH growth is accretion dominated even at z $\gtrsim$ 10. Moreover, only those BHs with weakened feedback prescriptions enter the "Merger Dominated" regime. None of the fiducial feedback BHs experience merger dominated growth in our (admittedly small) volumes. }
    \label{fig:GrowthFraction}
\end{figure*}

\subsection{Final Black Hole Masses}
\noindent The final masses of all BHs are shown in Figure \ref{fig:Mi_Mf} as a function of their initial masses at the final redshift of each region (i.e. independent of the growth mechanism be it through accretion or BH mergers or both) . The three feedback mechanisms are separated by different panels, the top panel shows the results from the NoFB runs, the middle panel the results from the WeakFB runs and the bottom panel the results from the FidBH runs. The different colors in each panel show results from the different zoom regions, blue for the \rarepeak region and red for the \NormalOne region.  The vertical gray bands mark the direct collapse channel for light seed BHs. Across all three feedback variants, we find that the distribution is largely independent of the region itself, despite being compared at different redshifts (\rarepeak at $z=15$ is similar to \NormalOne at $z=10$).

Crucially, we see that weaker feedback models ultimately allow  BHs to grow to higher final masses - the NoFB simulations have BHs growing larger than $10^8$ M$\odot$, while this maximum mass decreases to $10^7$ \msolar for WeakFB simulations, and down further to $10^6$ \msolar for the FidBH model.

Figure \ref{fig:Mi_Mf} also demonstrates that light seed black holes formed from PopIII stars are able to grow larger than $10^5$ M$_\odot$  in both the NoFB and WeakFB models, provided they transition into BHs without a SN explosion i.e. the stellar mass falls within the direct collapse mass range. This suggests that these BHs largely grow from their pre-natal gas clumps that would otherwise be destroyed through SNe. Although the resolution of this simulation suite of is state-of-the-art (1 physical pc spatial resolution and $10^4$ \msolar dark matter particle resolution) we are likely still not resolving accretion flows onto light seeds adequately and instead what we are seeing here is the tip of the iceberg in terms of light seed growth. Higher resolution simulations will be required to correctly resolve the small scale gas dynamics needed to track light seed growth \cite[e.g.][]{Mehta2024, Mehta2026}.

Furthermore, across all simulations, we find no strong correlation between the initial mass of the seed to the final BH mass. In particular, a larger initial seed does not necessarily evolve into a more massive BH. Seeds spanning approximately $10^3-10^5$ \msolar can ultimately grow to comparable masses of $\sim 10^7$ M$\odot$, while BHs with similar initial masses can follow substantially different growth trajectories. This indicates that the initial seed mass becomes progressively less important once BH growth in underway. Instead, the subsequent evolution is governed primarily by the local environment surrounding the BH. The availability and density of cold gas, the ability of the BH to retain access to dense gas, and its ability to sink to the galaxy's center all determine how rapidly a seed can grow. Thus, the final BH mass retains essentially no memory of its initial seed mass, but rather reflects the cumulative accretion history imposed by its environment.

\subsection{Black Hole Mergers}
\noindent We now investigate the BHs that underwent at least one merger, to quantify the contribution of mergers to their total mass growth.  As seen by Table \ref{tab:Simulations}, only a small fraction ($\sim5\%$ or less) of BHs have undergone mergers. We quantify the relative contributions of mergers and accretion in Figure \ref{fig:GrowthFraction}, which shows the relative growth of each BH from mergers and accretion on the y-axis versus the absolute value of growth on the x-axis.  The horizontal dashed line denotes equal growth from the two channels. BHs sitting above the dashed line have attained most of their mass through mergers (i.e. more than 50\% of the growth mass came from mergers). Conversely, BHs sitting below the line gained most of the mass through gas accretion. The different zoom regions are distinguished by marker style and the feedback prescriptions are denoted by colour. Among BHs that undergo mergers, accretion still remains the dominant growth channel for the majority of systems.  Only a small fraction ($\sim5\%$ of the growing BHs) obtain a larger fraction of their mass through mergers.\\
\indent Across all simulations, we identify two dominant types of mergers: mergers between two massive BHs (M1, M2 > $10^3$ M$_\odot$; hereafter called MM mergers), and mergers between a massive BH and a light BH  (M1 > $10^3$ M$_\odot$; M2 < 300 M$_\odot$, hereafter ML mergers).  The different mergers create distinct groups in the parameter space shown in Figure \ref{fig:GrowthFraction}. The BHs beneath the dashed-dot line all underwent a single ML or multiple ML events, gaining less than 300 \msolar through mergers. Their growth remains largely accretion dominated.  All BHs lying above this line have gained considerable mass through mergers, and have undergone at least a single MM merger.

The FidBH runs exhibit the smallest number of mergers, with only $\sim 40$ BHs undergoing mergers in each zoom region. In these simulations, the majority of heavy seed BHs reach final masses between $10^4-10^5$ \msolarc, with a small tail extending beyond $10^6$ \msolarc. There are no BHs in the FidBH simulations that gain more mass through mergers than through accretion, i.e, they all lie beneath the dashed lines. 

For the WeakFB and NoFB simulations, we have considerably more BHs undergoing mergers,  $\sim 800$ for NoFB and $\sim 400$ for WeakFB. While the majority of BHs smaller than $10^7$ \msolar are accretion dominated, mergers become essential for assembling the most massive BHs. Extreme BH masses ($>10^8$ M$_\odot$) are only achieved in the NoFB simulations, where the number of BHs undergoing mergers also increases substantially. For example, 55 ($\sim0.5\%$) and 56 ($\sim 1.5 \%$) BHs in the RP-NoFB and N1-NoFB simulations, respectively, gain more mass through mergers than through accretion.

\begin{figure}
    \centering
    \includegraphics[width=\linewidth]{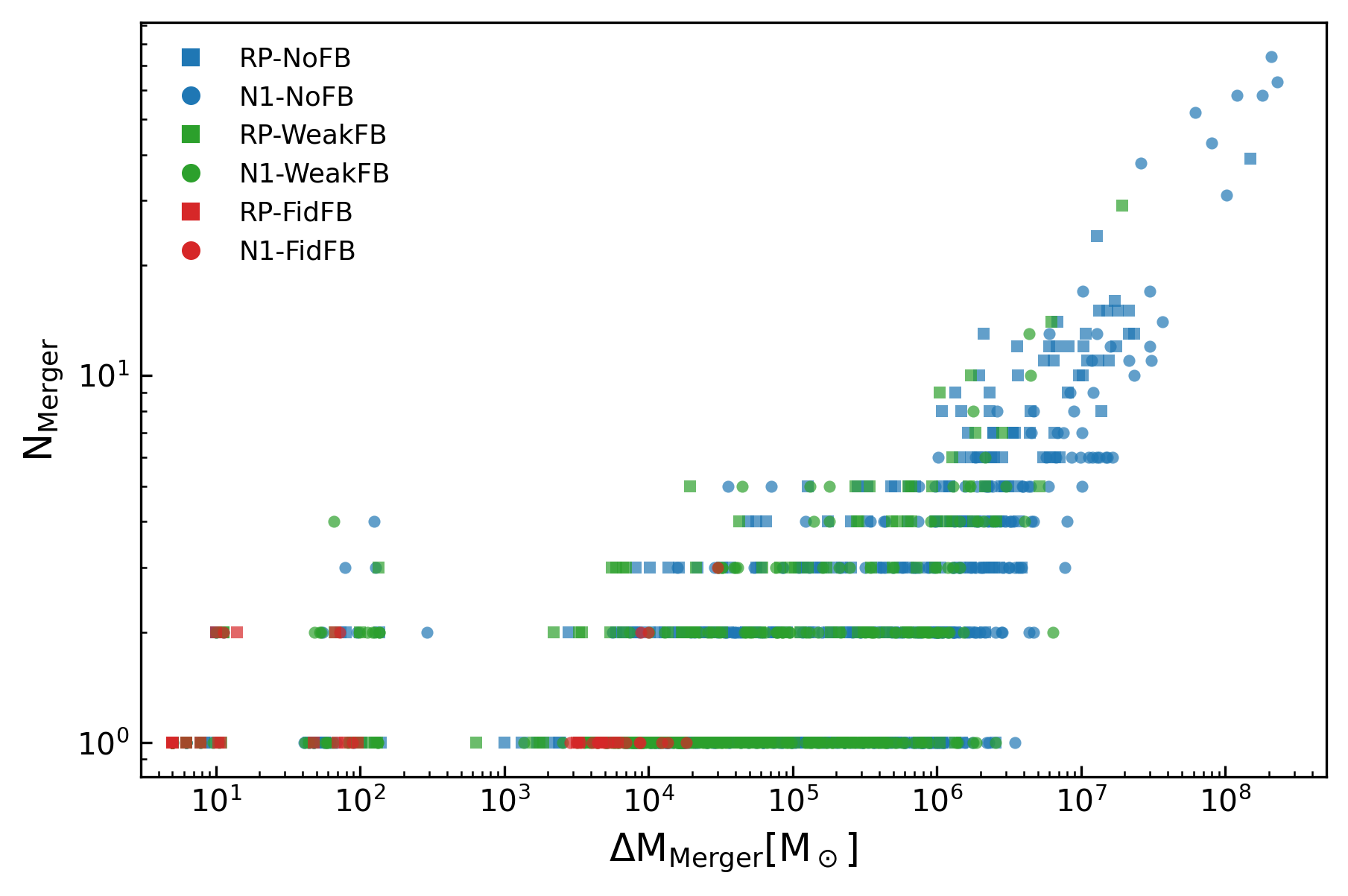}
    \caption{Number of mergers a BH undergoes as a function of mass it gains through mergers. The different region are shown by different marker styles and feedback prescriptions have different colours (blue for NoFB, green for WeakFB and red for FiducialFB). What is clear is the number of mergers and the mass gained through mergers is highest in the no feedback simulations. }
    \label{fig:N_merger}
\end{figure}

\subsection{Tracing Black Hole Merger Histories}
\noindent We next investigate whether merger-driven BH growth is dominated by a single major merger event or by repeated hierarchical mergers. In Figure \ref{fig:N_merger}, we show the number of mergers experienced by each BH as a function of the total mass gained through mergers. Across all simulations, BHs that undergo only a single merger can acquire anywhere between $\sim 5 $ and $5\times10^6$ \msolar through that event alone. Conversely, some BHs gain only $\sim 10^2$ \msolar through mergers despite experiencing up to four merger events. This demonstrates that the cumulative merger contribution depends not only on the number of mergers, but also on the masses of the merging BHs.

A clear dependence on feedback strength is also evident. BHs in the FidBH simulations undergo at most three mergers, indicating that strong feedback efficiently suppresses repeated hierarchical BH assembly. In contrast, the maximum number of mergers experienced by a single BH rises to $\sim 30$ and $\sim 70$ in the WeakFB and NoFB simulations, respectively. Despite these extreme merger histories, the majority of BHs across all simulations still experience fewer  than five mergers. Both MM and ML merger channels contribute to these highly hierarchical growth pathways.

\begin{figure*}
    \centering
    \includegraphics[width=0.8\linewidth]{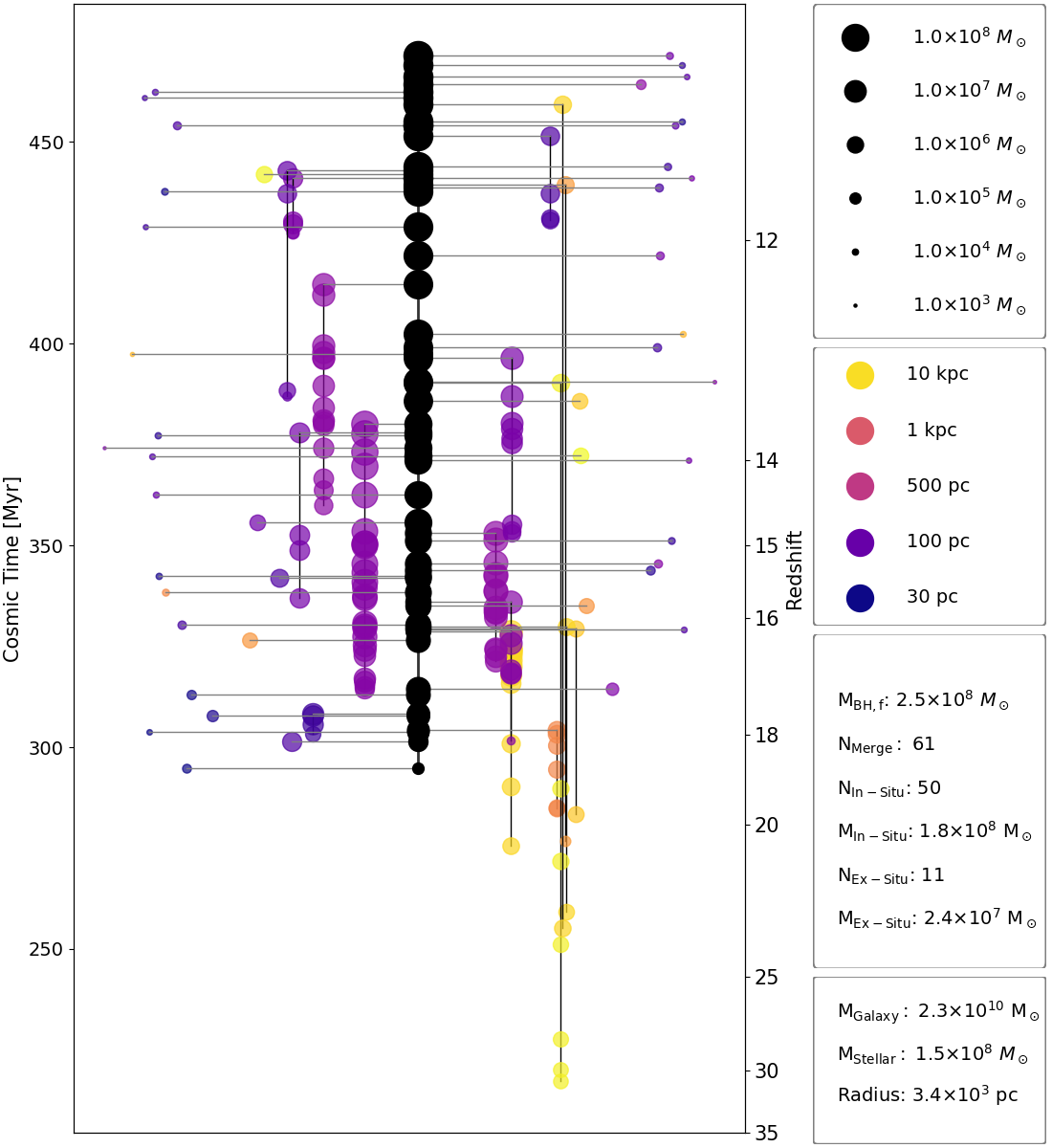}
    \caption{Merger tree for the most merging BH in the N1-NoFB simulation. The central black line tracks the primary BH, with each circle denoting a merger event. Each circle is connected through a horizontal line to the secondary BH. For each secondary BH, we also trace its own merger history through vertical lines and circles. We do not show here tertiary BHs merging into the secondary ones. The circles are scaled according to the BH mass at those times. The color scheme shows how far the secondary BH was at its time of formation from the primary BH.}
    \label{fig:MergerTree_NoFB}
\end{figure*}

The most actively merging BH in the N1-NoFB simulation gains more than $2.5 \times 10^8$ \msolar through mergers alone and undergoes a total of 75 merger events (14 ML mergers and 61 MM mergers). This BH lies in a galaxy with a halo mass $\sim 2.33 \times10^{10}$ \msolar and a stellar mass of $\sim 2.08\times10^8$ \msolarc, clearly having an overmassive BH compared to its stellar component. We investigate its hierarchical assembly history in Figure \ref{fig:MergerTree_NoFB}. The central black line traces the growth history of the primary BH. Each circle denotes a MM merger event involving a secondary BH with mass $M_2 > 10^3$ \msolarc. The size of each circle is scaled according to the BH mass at the time of merger.

Horizontal lines connect the primary BH to the merging secondary BHs. For each secondary BH, the accompanying vertical branch shows its own merger-driven growth history prior to merging with the primary BH. While we do not explicitly show tertiary BHs merging into the secondary BHs, the branching structure indicates how many merger events each secondary BH experienced before joining the primary merger tree. The secondary BH branches are additionally color-coded according to their initial separation from the primary BH at the time of formation. This allows us to distinguish between BHs that likely formed within the same host galaxy and those introduced through galaxy assembly and subsequent dynamical evolution. The halfmass-radius for this galaxy is around $3.4$ kpc.

The majority of secondary BHs, with masses below $10^6$ \msolarc, form and subsequently merge with the primary BH without experiencing any prior merger events. These BHs typically span masses between $10^3$ and $10^6$ \msolarc. In contrast, secondary BHs more massive than $10^6$ \msolar typically exhibit their own hierarchical merger histories prior to merging with the primary BH. The most massive secondary BH reaches a mass of $\sim 8 \times 10^7$ \msolar and undergoes approximately 25 mergers before joining the primary merger tree.

Approximately $75\%$ of the secondary BHs form within $\lesssim 3.4$ kpc of the primary BH, corresponding to the blue and purple branches in Figure \ref{fig:MergerTree_NoFB}. These BHs likely originate within the same host galaxy and subsequently migrate towards the galactic center through dynamical interactions and nuclear clustering, contributing a total of $\sim 1.8\times10^8$ \msolar through mergers. The remaining BHs form at larger separations and are likely introduced through hierarchical galaxy assembly and subsequent galaxy mergers. Despite being fewer in number, these externally formed BHs still contribute an addition $\sim 2.8\times10^7$ \msolar to the final BH mass.

 \begin{figure*}
     \centering
     \includegraphics[width=0.8\linewidth]{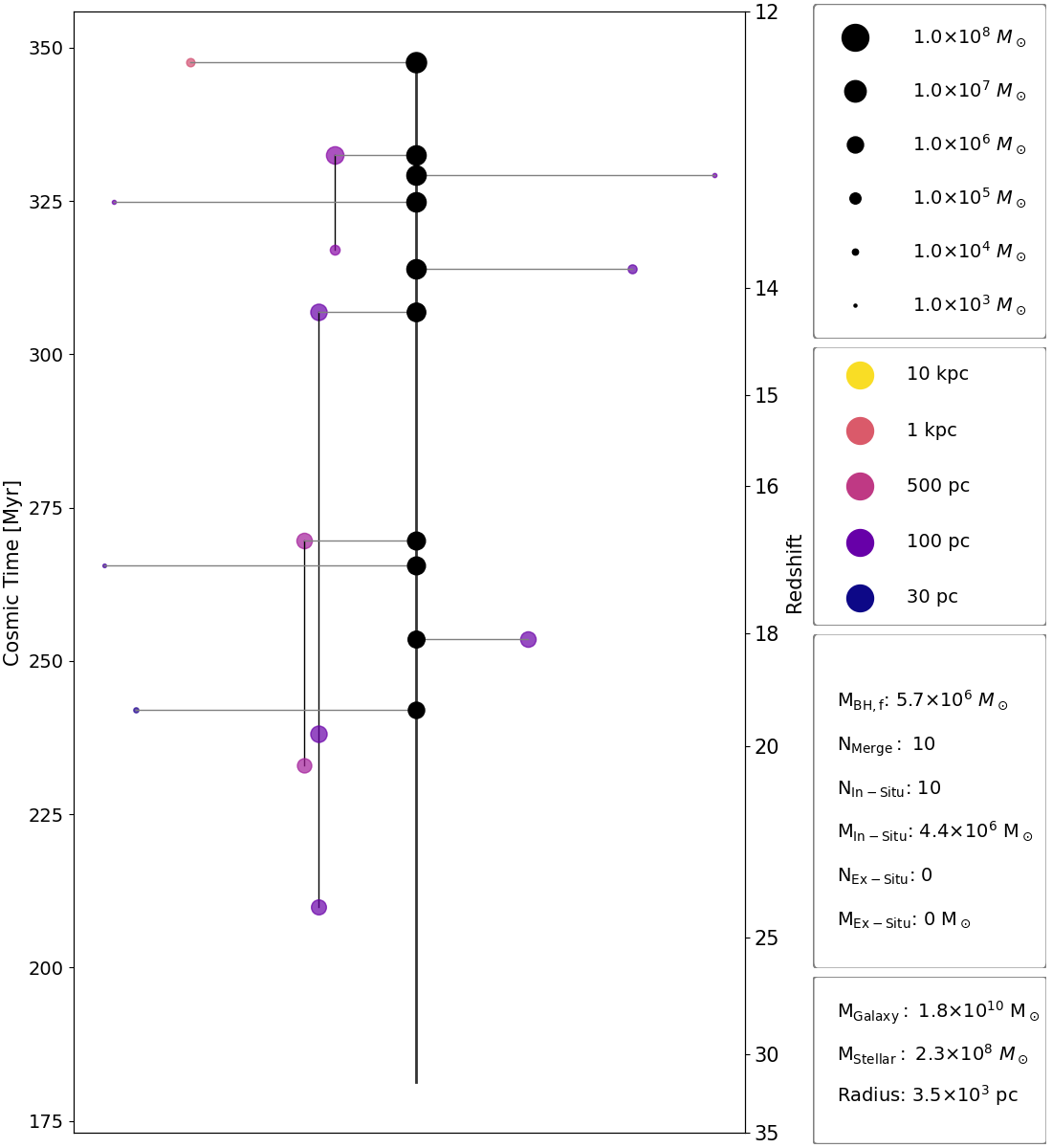}
     \caption{Same as Figure \ref{fig:MergerTree_NoFB} for the most merging BH in N1-WeakFB simulation.}
     \label{fig:MergerTreeWeakFB}
 \end{figure*}
We similarly study the most actively merging BH in the N1-WeakFB simulation to further examine the impact of feedback on hierarchical assembly of these BHs. This BH gains more than $4.3\times10^6$ \msolar after undergoing 13 mergers (10 MM and 3 ML). The host halo has a mass of $\sim 1.8\times10^{10}$ \msolar with a stellar mass of $\sim2.3\times10^8$ \msolarc. This is an equivalent galaxy to the previous case but the BH gained almost two orders of magnitude less mass than the NoFB case. We see that in Figure \ref{fig:MergerTreeWeakFB}, the largest secondary BH to merge is only $\sim 1.74\times10^6$ \msolarc, and has only undergone 1 merger of its own before merging into the primary. Moreover, all secondary BHs were formed within 500 pc of the primary, suggesting that it only gained mass through in-situ mergers. 

In contrast to the N1-NoFB simulation, the most merging BH in the N1-FidBH run undergoes only a single major merger, resulting in a merger tree consisting of a single branch connecting two BHs. This galaxy is also similar in properties with the previous two galaxies with host halo mass of $2.2\times 10^{10}$ \msolar and stellar mass of $4.1\times10^8$ \msolarc.

\subsection{BH Merger Suppression due to Feedback}

To understand the origin of this dramatic suppression of hierarchical assembly, we investigate several properties of the host galaxies across different feedback models. We find that the total gas content of galaxies remain broadly similar between the NoFB, WeakFB, and FidBH runs. Likewise, the number of BH seeds formed, the BH occupation fractions, and the overall galaxy growth histories show only minor differences across the simulations (with different feedback prescriptions). These results suggest that the suppression of BH mergers is not driven by changes in galaxy assembly, seed formation, or the global gas reservoir available within galaxies.

\begin{figure}
    \centering
    \includegraphics[width=\linewidth]{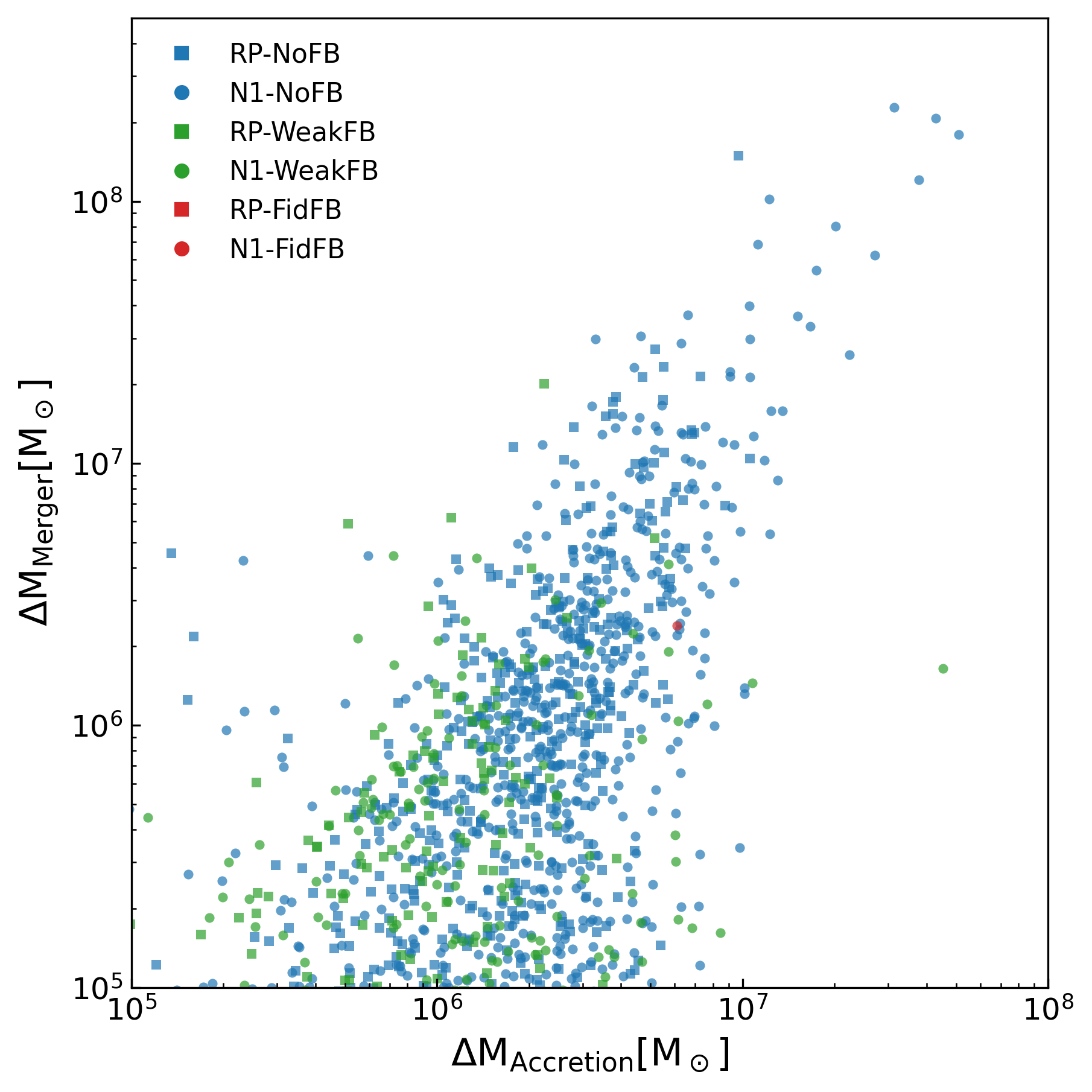}
    \caption{Mass gained by a BH through mergers as a function of mass gained through accretion. The different region are shown by different marker styles and feedback prescriptions have different colors. Growth achieved through mergers is tightly coupled to growth through accretion - particularly at the higher mass end. What we find is that BHs must initially grow by accretion until reaching a tipping point of approximately $10^6$ \msolar (shown here by the tightening of the scatter) after which growth through BH-BH mergers can be attained. }
    \label{fig:AccretionMerger}
\end{figure}

We therefore conclude that the dominant effect of accretion feedback operates on much smaller scales. In Figure \ref{fig:AccretionMerger}, we show the total mass gained through mergers as a function of the mass gained through accretion.  At high masses, we find a clear positive correlation, indicating that BHs experiencing substantial merger-driven growth have also typically undergone significant accretion growth. This suggests that early accretion is an important prerequisite for entering highly hierarchical merger pathways. In the FidBH simulations, accretion feedback disrupts the dense gas clumps surrounding newly formed BHs, prematurely terminating their initial growth phase. As a result, BHs remain close to their seed masses for significantly longer periods of time than in the NoFB and WeakFB runs.  This divergence in BH accretion growth naturally reduces the likelihood that BHs in the FidBH simulations become dominant members of the nuclear BH population and participate in repeated hierarchical mergers. 

Considerable scatter is present at lower masses in Figure \ref{fig:AccretionMerger}, indicating that accretion growth alone does not uniquely determine the subsequent merger history of a BH. Additional environmental and dynamical factors must also contribute. Taken together, these results suggest that feedback suppresses hierarchical BH assembly primarily by regulating the local gaseous environments that govern the earliest stages of BH growth, while secondary processes determine the detailed merger histories of individual systems.

\section{Gravitational Wave Detection Predictions} \label{Sec:GWDetections}

\noindent We have shown that the implementation of BH feedback has a strong impact on the hierarchical growth of BHs through mergers. Massive BH mergers provide a complementary observational window into the assembly of massive BHs through gravitational-wave astronomy. Future observatories such as the Laser Interferometer Space Antenna (LISA; \citealt{Amaro-Seoane2023}) and potentially the Einstein Telescope (ET; \citealt{Puntaro2014}), and the Lunar Gravitational-Wave Antenna (LGWA; \citealt{Harms2021}) will probe different portions of the BH merger population. LISA is particularly well suited to observing massive BH binaries (chirp masses between approximately $10^4$ \msolar and a few times $10^7$ \msolarc), while LGWA will probe an intermediate-frequency regime (chirp masses between approximately $10^3$ \msolar and a few times $10^5$ \msolarc) and can provide complementary observations of lower-mass systems.

We therefore extend our merger catalog to estimate which of the events in our simulations could be detectable by these future observatories. Although ET is sensitive to lower-mass BH binaries than LISA, only a small number of mergers in our simulations fall within the mass range ET has appreciable sensitivity at the redshifts considered here. In our simulated population, the events with chirp masses below ($10^3$ \msolarc) do not produce signal-to-noise ratios (SNRs) above our adopted detection threshold. We therefore focus the remainder of this analysis on LISA and LGWA. 

For each merger, we calculate the chirp mass,
\begin{equation}
    M_{chirp} = \frac{(M_1M_2)^{3/5}}{(M_1 + M_2)^{1/5}}.
\end{equation}
where $M_1$ and $M_2$ are the primary and secondary BH masses, respectively. The frequency of the GW emitted is directly correlated to the chirp mass. We then use the component masses and merger redshift to calculate the expected SNR using \gwfish \citep{Dupletsa_2023}. The GW signal is modeled using the inspiral-merger-ringdown \texttt{IMRPhenomD} \citep{Purrer} waveform model with quadrupole only dominant mode (l, m) = (2, 2). The remaining source parameters such as the sky localization ($ra = 0.0$, $dec=0.0$), coalescence phase ($\phi_c = 0$), time of merger in geocentric coordinates ($t_c = 10^9$ yrs) are kept as default, so the absolute numbers should be interpreted as forecasts under these assumptions. We further adopt an SNR threshold of SNR $> 8$ to define a potentially detectable event. 

\begin{figure*}
    \centering
    \includegraphics[width=\linewidth]{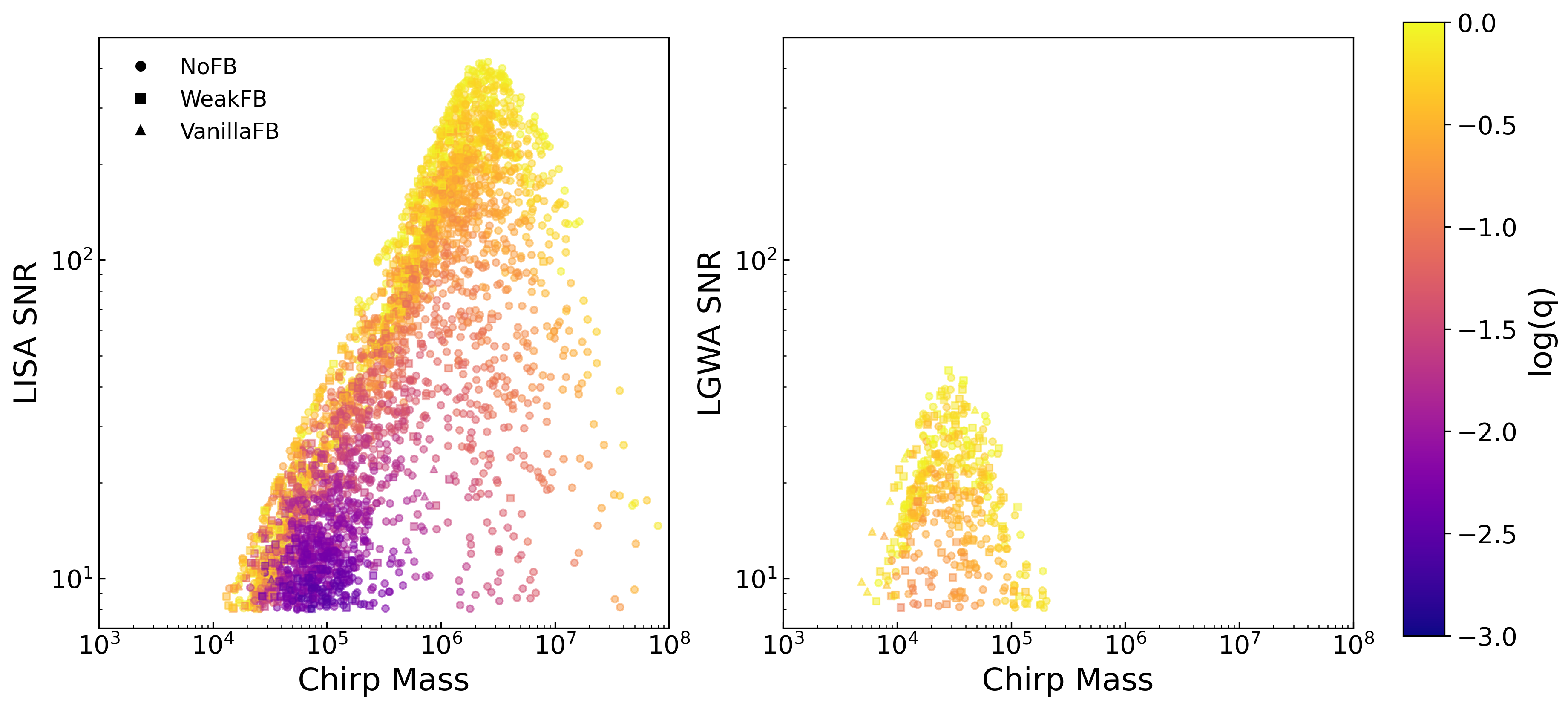}
    \caption{SNR vs chirp mass of BH merger from all the simulations. The dots are further colored according to their mass ratios, q. In the left hand panel we plot the SNR values for LISA versus the chirp mass. LISA has peak sensitivity at chirp masses of approximately $10^6$ \msolar with near equal mass mergers. In the right hand panel we show the same plot for the LGWA which probes smaller (chirp-)mass mergers. In the case of LGWA the instrument is most sensitive to chirp masses of approximately $5 \times 10^4$ \msolar and again near equal mass mergers.\\}
    \label{fig:SNR}
\end{figure*}

Figure \ref{fig:SNR} shows the predicted SNR as a function of chirp mass for all mergers in our simulations. The points are colored according to the mass ratio ($\rm{q = M_2/M_1}$), while different marker styles denote the adopted BH feedback prescription. The left and right panels show the predicted SNR for LISA and LGWA, respectively.

For LISA, the detectability exhibits a strong dependence on chirp mass. The SNR increases rapidly with chirp mass, reaching its highest values for systems with $\rm{M_{chirp}\sim 10^6-10^7}$ \msolar (this corresponds to the LISA frequency bucket - i.e. where LISA is most sensitive), where the strongest events reach SNRs of several hundred. At still higher masses, the SNR decreases as the characteristic gravitational-wave frequencies of the binaries move beyond the most sensitive part of the LISA frequency band. Conversely, lower-mass systems produce weaker signals, resulting in a broad lower-SNR population. The resulting envelope therefore reflects both the increasing gravitational-wave amplitude with mass and finite frequency response of the detector.

The mass ratio provides a second important dependence. At fixed chirp mass, systems with more equal component masses generally produce stronger signals than highly unequal-mass binaries. This is evident in the figure, where the highest SNR events are predominately systems with q $\sim 1$, while mergers with q $\lesssim10^{-2}$ preferentially occupy the lower-SNR portion of the distribution. Nevertheless, a substantial population of unequal-mass mergers remains detectable, demonstrating that LISA is sensitive to a broad range of merger configurations predicted by our simulations.

LGWA probes a distinctly different portion of the merger population. The highest SNRs occur over a narrower chirp mass range of approximately $10^4-10^5$ \msolar with near equal-mass systems reaching SNRs of almost 40. The SNR decreases rapidly toward lower mass ratio, with systems having q $\lesssim0.1$ generally falling below our detection threshold. Thus, while LISA preferentially detects the more massive mergers in our population, LGWA provides complementary sensitivity to lower-mass systems. This demonstrates the value of considering multiple GW observatories when interpreting the merger population of the early Universe.

\begin{figure*}
    \centering
    \includegraphics[width=\linewidth]{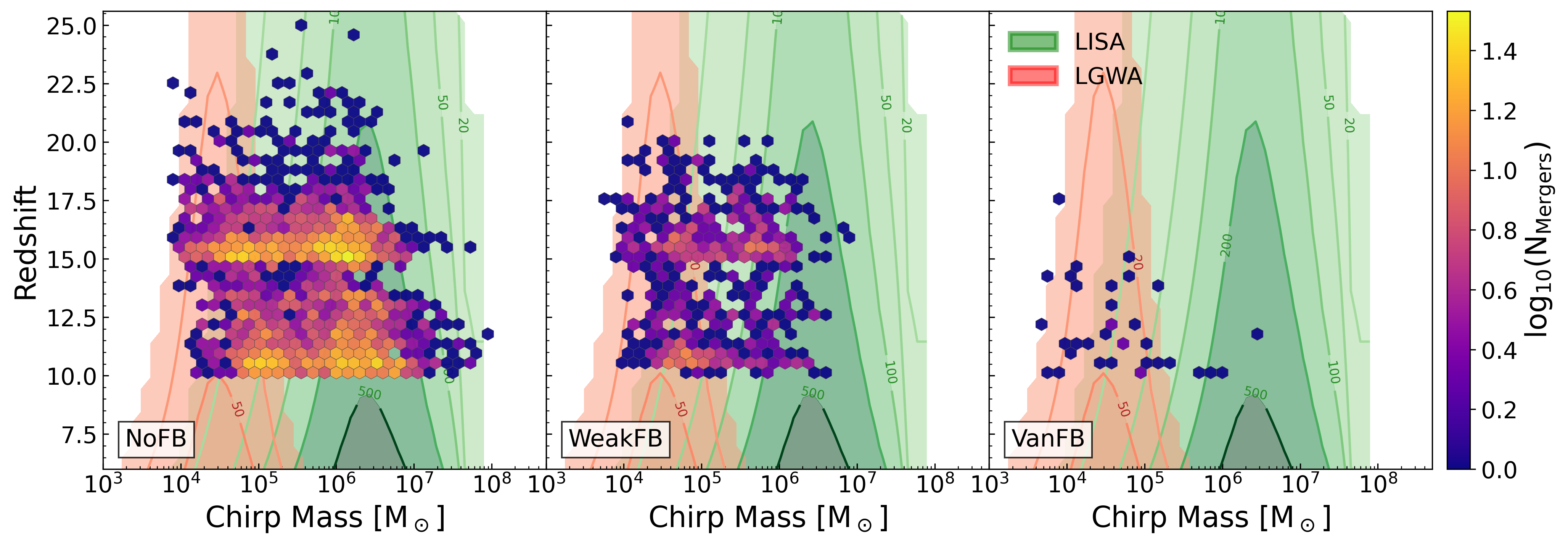}
    \caption{Chirp mass versus redshift for each of our feedback models. In green we show the waterfall contours for LISA while in orange we show the waterfall contours for LGWA. The hexbins show mergers from the simulations. For the NoFB cases we see, as expected, the highest number of mergers. We also see two distinct population for \rarepeak and \NormalOne simulations (most visible in the left and middle panels). The number of massive BHs and hence the number of massive BH mergers reduces as a function of feedback strength. For the Full Feedback case the number of mergers has dropped by a factor of 100 compared to the NoFB case. Nonetheless, mergers in our simulations are potentially detectable by both LISA and LGWA.}
    \label{fig:chirpmass}
\end{figure*}
We also notice that the feedback prescriptions produce broadly similar merger properties but strongly affect the abundance of BH mergers. We now investigate whether these differences in abundance could be distinguished through future GW observations. Figure \ref{fig:chirpmass} shows the distributions of merger chirp mass and redshift for three feedback prescriptions, combing the \rarepeak and \NormalOne regions. The background contours indicate the expected SNR of LISA (green) and LGWA (red), allowing the regions of parameter space accessible to each detector to be identified directly. The distributions are broadly similar in their overall extent, but their populations differ substantially in number. 

The NoFB simulations produce the largest merger population, spanning almost four orders of magnitude in chirp mass, from $\sim 10^4$ to $\gtrsim10^8$ \msolarc, and extending over the full redshift range probed by our simulations, $10 \lesssim z\lesssim25$. Two concentrations are visible in the distribution, corresponding primarily to the different redshift ranges reached by the \rarepeak and \NormalOne simulations. The NoFB population also extends furthest into the high chirp-mass regime, producing mergers with $\rm{M_{chirp} \gtrsim10^7}$ \msolar that are largely absent in the feedback-regulated models.

Introducing feedback progressively reduces the number of mergers occupying this parameter space. The WeakFB simulations retain a substantial merger population, but with far fewer events than the NoFB case and with a reduced high-mass tail. The suppression is even stronger in the FidBH simulations, where only a small number of mergers remain and none reached the highest chirp masses found in the NoFB population. Thus, feedback does not appear to introduce a fundamentally new class of merger systems; rather, it suppresses the frequency with which the systems capable of undergoing hierarchical mergers are produced.

The effect on the observable population can be quantified by counting the mergers that fall above the adopted SNR detection threshold. These numbers are summarized in Table \ref{tab:GW-detections}. Across the two regions, the NoFB simulations produce 2613 potentially detectable LISA events, compared to 603 for WeakFB and only 24 for FidBH. Thus, the number of predicted LISA detections decreases by nearly a factor of five from NoFB to WeakFB and by almost two orders of magnitude from NoFB to FidBH. The same trend is present for LGWA, although the absolute number of detectable events is smaller: 284, 139, and 16 events predicted for NoFB, WeakFB, and FidFB, respectively. 

We identify 196, 93, and 5 events, respectively that satisfy the detection threshold for both LISA and LGWA. Such multi-band detections would provide complementary information on the binary parameters and could substantially improve parameter estimation. More importantly for this work, combining observations across different GW frequency bands could help reconstruct the mass and merger distribution of early BHs over a broader range of masses.

\begin{table}[]
    \centering
    \begin{tabular}{c|c|c|c} \toprule
         Feedback&   LISA SNR $>8$& LGWA SNR $> 8$&LISA + LGWA\\
         \midrule
         NoFB& 2613 & 284 & 196\\
         WeakFB& 603 & 139 & 93\\
         FidBH& 24 & 16 & 5\\
        \bottomrule\end{tabular}
    \caption{The absolute number of LISA and LGWA events in our SEEDZ volumes. The first column is for LISA only, the second column is for LGWA and the final column counts the number of detectable events that would be detectable by both detectors. We apply a SNR threshold of 8 in all cases. }
    \label{tab:GW-detections}
\end{table}

The predicted detectability also provides an observational connection to our feedback results. Since the feedback prescriptions primarily change the abundance of mergers rather than producing dramatically different distributions of merger masses, the principal observable signature of feedback is expected to be a difference in the number of detectable events. A sufficiently large sample of high-redshift GW detections could therefore provide an independent probe of the physical processes regulating BH growth and hierarchical assembly in the early Universe.

\begin{figure}
    \centering
    \includegraphics[width=\linewidth]{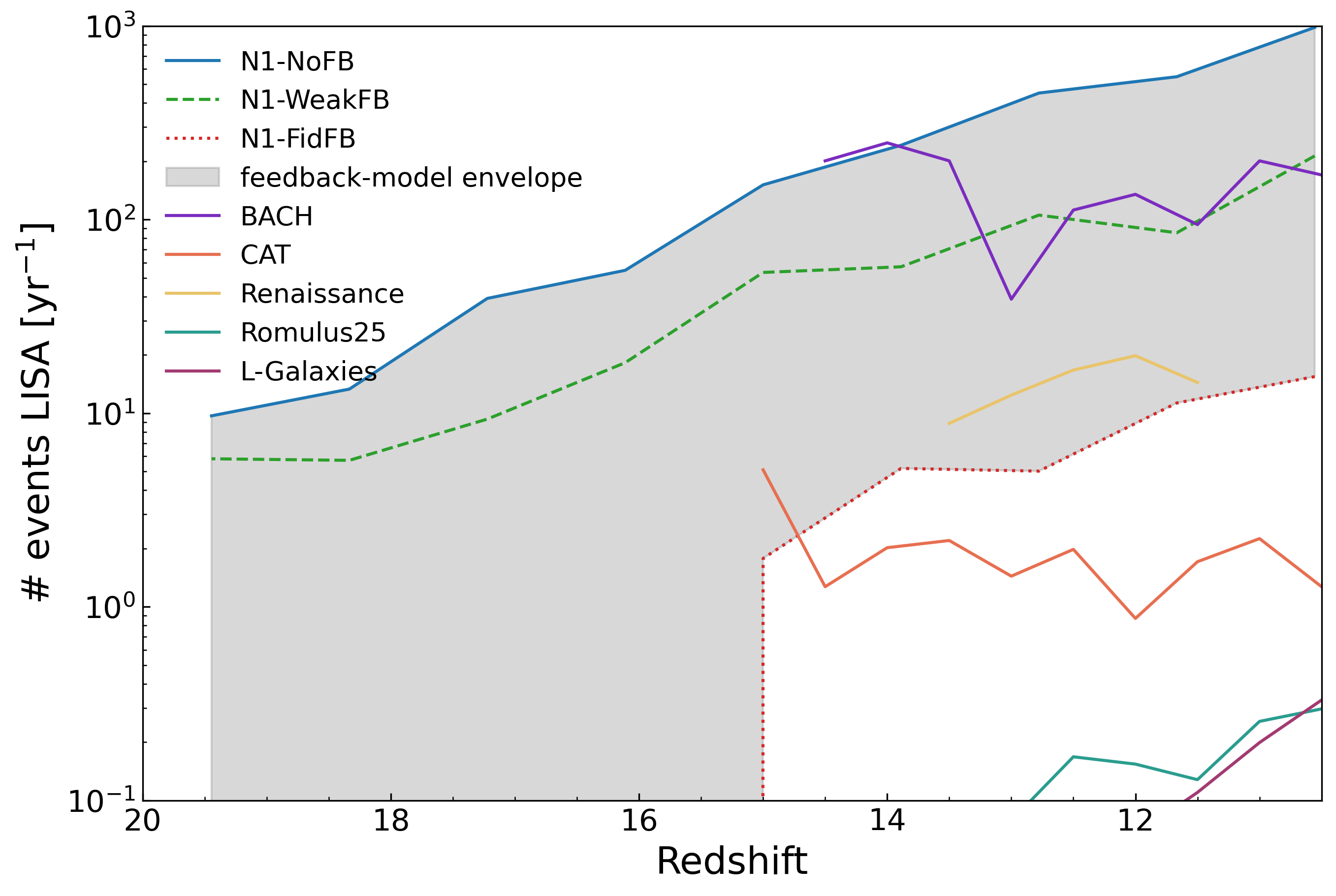}
    \caption{Number of LISA detectable events per year as a function of redshift. Here we show the number of detections per year that LISA may be expected to see as a function redshift. The blue line is the noFB case, the green dashed line is for the WeakFB case and the red dotted line for the FiducialFB case. The grey shaded region therefore covers the full parameter space. From the SEEDZ model LISA may see up to a hundred mergers per year in a weak feedback scenario. Our results indicate a strong degeneracy between the seeding process (fixed here) and the feedback mechanism (varied here). Even with ubiquitous seeding, feedback depending on its efficiency may suppress merger events. Comparing to other analagous models SEEDZ is consistent with other high redshift models including BACH and Renaissance while lying somewhat above CAT, L-Galaxies and Romulus25.}
    \label{fig:lisa_events}
\end{figure}

We further quantify the redshift dependence on the observable merger population by estimating the number of LISA-detectable events per year in successive redshift intervals (Figure \ref{fig:lisa_events}). For this analysis, we focus on the \NormalOne region, which provides a representative environment relative to the overdense \rarepeak region and is evolved to lower redshifts. The resulting distribution spans approximately $10\lesssim z \lesssim20$, allowing us to examine how the predicted observable merger population evolves. We account for the limited volume of the zoom-in simulations by applying a comoving-volume normalization to the measured merger counts. The resulting values represent the merger abundance expected when scaling the simulated population to the volume of the parent region.

The predicted number of detectable events increases systematically towards lower redshifts in all feedback models. This reflects the increasing abundance of massive BHs and mergers as structure formation progresses, providing LISA with an increasingly large population of sources at later cosmic times. However, the normalization of this population depends strongly on the adopted feedback prescription. The NoFB model provides the upper envelope, with the number of potentially detectable events increasing from only a few events per year at z $\sim 19$ to almost a thousand events per year by z $\sim 10$. The WeakFB model lies systematically below this prediction, reaching a hundred events per year at z $\sim 10$. In contrast, the FidBH model produces the lowest event rate, remaining at roughly $\lesssim10$ events per year event at z $\sim 10$.

The resulting spread between the NoFB and FidBH predictions defines a broad feedback-dependent envelope for the expected high-redshift LISA event population. At z $\sim 10$, our models span approximately two orders of magnitude in the predicted number of detectable events. This provides a potentially powerful observational diagnostic: rather than requiring individual mergers to uniquely identify the underlying BH growth model, a sufficiently large sample of high-redshift LISA events could constrain the efficiency of BH feedback through the evolution of the merger rate with redshift.

We also compare our LISA detectable rates with others in the literature \citep{Villalba2026}, only addressing the models that incorporate mergers at redshifts higher than 10.  We find most of the models predict a lower event rate than our simulations. Only BACH \citep{Barausse2023} and Renaissance \citep{McCaffrey2024} models have event rates within our envelope. BACH  in particular match our WeakFB results well as they include a very efficient heavy-seed model scenario. Renaissance has similar resolution to \seedz and gives similar results. 

\section{Discussion}
\label{sec:discussion}

In this work, we have shown BH assembly through hierarchical mergers in the high-resolution SEEDZ simulations. We analyze a total of six simulations of two regions \rarepeak and \NormalOne with three variants of feedback NoFB, WeakFB, and FidBH. We showed that for the assembly of the largest BHs, mergers are an important requirement. We also showed how different BH feedback prescriptions altered the growth of BHs through mergers, which is not trivial or straightforwards.

Our results can be compared with the recent BRAHMA \citep{Bhowmick2024, Bhowmick2024a, Bhowmick2025b} simulations, which also investigates the role of BH mergers in the assembly of the first massive BHs. \citet{Bhowmick2025b} found that mergers can dominate the early growth of low-mass BH seeds, with gas accretion contributing less than $\sim 2\%$ of the final mass of their merging BH population. Their simulations predict merger rates of $\sim 100-1000 ~\rm{yr}^{-1}$ at z $\gtrsim5$, depending on the adopted seed model. In contrast, we find that accretion dominates the growth of the majority of BHs in our simulations, while mergers become increasingly important for the most massive BHs. this difference does not necessarily imply conflicting physical pictures since we are probing different scales and physical processes.

Several methodological differences between the two simulation suites may also contribute to the contrasting relative importance of accretion and mergers. First, the criteria adopted for BH mergers are substantially different. In BRAHMA, BHs are allowed to merger once they approach a minimum separation of 90 pc, wheres our merger criterion required BHs to reach separations of only $\sim 5$ pc. Our treatment therefore imposes a considerably more restrictive condition on the final stage of BH coalescence and is likely to suppress mergers that would be counted in BRAHMA at larger separations. 

The accretion prescriptions also differ in an important way. The BRAHMA simulations employ Eddington-limited accretion, whereas our SEEDZ model permits super-Eddington accretion under the dense gas conditions commonly encountered around newly formed BHs. Consequently, BHs in our simulations can undergo rapid early growth that is not available to a strictly Eddington-limited population. This distinction can alter the relative importance of the two growth channels: rapid accretion allows initially small seeds to reach substantially larger masses before they participate in subsequent mergers, whereas limiting accretion naturally increases the relative contribution of mergers to the growth of the BH population. The different accretion prescriptions therefore provide a plausible explanations for why mergers contribute a much larger fraction of the BH growth in BRAHMA than in our simulations.

Another important distinction comes from their use of a large dynamical mass when computing  the dynamical friction drag on black holes. This might cause black holes to sink efficiently to the center of galaxies, which we do not observe in our simulations.

Finally, the two studies probe different evolutionary timescales. The merger timescales reported in BRAHMA are typically of order of a few hundred to a few thousand Myr, whereas our simulations follow the early Universe to only cosmic times of approximately 400 Myr after the Big Bang. Many of the BHs that we identify as potential merger participants therefore have not had sufficient cosmic time to complete the long dynamical evolution required to coalescence. Our results should consequently be interpreted primarily as describing the early stages of hierarchical BH assembly, rather than the eventual merger history of these BH populations. The substantially longer evolution timescales accessible to BRAHMA allow mergers that would occur well beyond the endpoint of our simulations to contribute to their inferred BH growth. This temporal difference is particularly important when comparing merger rates and the fraction of BH mass assembled through mergers between the two suites.

We also compare our results with the FLARES simulation \citep{Liao2025}, who focused on supermassive BH mergers between a redshift of $5 \lesssim z \lesssim 10$. They use the dark matter seeding method where they seed a halo with a BH of $10^5$ \msolar once it satisfies a given halo mass condition. One of the major results of their work is that overdense region seems to have a significant impact on the final masses of the BH and thus on BH merger. This does not contradict our results since our \rarepeak simulation end at z $=15$. If we continue the simulation to a redshift of 10, we too would see the impact of overdense region on BH mergers. They also show a LISA detectable rate of $0.03 \rm{yr}^{-1}$, which is two of three orders of magnitude lower than ours a result likely dependent on the very different seeding prescriptions employed by \seedz~ and FLARES. 

\citet{McCaffrey2024} also performed a similar analysis to that outlined here. They use the Renaissance simulations \citep{Xu2016, Chen2014a, OShea2015,Wise2019, Regan2020, Regan2020a} to post-process the formation and growth of heavy seeds in the Renaissance datasets. They similarly find that massive BHs grow primarily through accretion rather than mergers. \citet{McCaffrey2024} show that during its expected mission lifetime (4 years), LISA may detect up to 80 merger events from Normal1 region and a 1000 events from Rarepeak region. These predicted detection rates are broadly consistent with the detection rates we predict here.

Lastly, we compare our results to \citet{Davari2026} who performed a complimentary analysis to ours using the Cosmic Archaeological Tool (CAT; \citealt{Trinca2022}). CAT has models for light seeds, heavy seeds, and intermediate seeds, similar to what was used in this study. Using CAT, they performed an analysis on binary BH mergers from a redshift range of 24 to 4, using multiple model for BH accretion (Eddington-Limited and Super-Eddington). They also used \gwfish to calculate the SNR for gravitational wave detections for LISA, ET, and LGWA. And they found that the merger rates for events depend largely on the accretion model selected with Super-Eddington model allowing for more mergers. This is analogous to our results where a massive BH population allows for more mergers. However, comparing our results, we find a degeneracy between different accretion and feedback models that would need further investigation. There is also a difference in the number of merger events detected by LISA by almost 2 order of magnitude from our median weak feedback simulations. This apparent difference could originate from their inclusion of a dynamical merger timescales while we merge our BH immediately.

\section{Conclusions}
\label{sec:conclusions}

In this work, we have investigated the growth and merger history of BHs in the early Universe using the high-resolution SEEDZ simulations. We followed BH populations in two zoom regions with different BH feedback prescriptions, allowing us to isolate the role of feedback in regulating both accretion and merger-driven growth. We further connected the resulting merger population to the expected observations of future gravitational-wave observatories focusing on LISA and the LGWA.

Our main conclusions are:
\begin{itemize}
    \item Early black hole growth, starting from the seed mass, is exclusively accretion dominated. At masses in excess of approximately $10^6$\msolar we see a turnover from accretion dominated growth to merger dominated growth. Once the BH mass exceeds approximately $10^7$ \msolar BH growth becomes merger dominated in our models

    \item BH mergers can proceed hierarchically in the early Universe (once they becomes sufficiently massive). In the no feedback simulations, individual BHs can undergo more than 20 successive mergers, with the most extreme case undergoing 74 mergers and gaining the final $> 10^8$ \msolar through mergers alone. These mergers arise from both BHs formed within the same galaxy and BHs brought together through galaxy mergers.
    
    \item BH feedback strongly suppresses hierarchical merger growth. The merger population changes dramatically with the strength of accretion feedback. While the NoFB simulations produce BHs undergoing tens of repeated mergers, simulations including thermal feedback due to BH accretion typically produce no more than a few mergers per BH. Importantly, this suppression is not primarily driven by differences in the large-scale galaxy population: the number of seeds, galaxy masses, star formation, gas content, and galaxy merger histories remain broadly similar across the feedback variants. 
    
    \item BH Feedback regulates mergers indirectly through the early growth of BHs. Our results indicate that the dominant effect of BH feedback occurs on the local scales surrounding newly formed BHs. Accretion feedback disrupts dense gas structures, suppressing the initial growth of BHs thus leaving the BHs less massive (compared to our no feedback models).  These less massive BHs are less able to become dynamically dominant within the nuclear BH population and consequently participate in fewer hierarchical mergers. The correlation between accretion-driven and merger-driven growth further supports this picture: BHs that gain substantial mass through mergers generally experienced substantial accretion growth.
    
    \item The merger population provides a potential gravitational-wave probe of early-Universe feedback. Although the different feedback models occupy broadly overlapping regions of the chirpmass-redshift parameter space, they produce dramatically different number of mergers. In our predictions, the number of potentially detectable LISA events changes from thousands per year in the no feedback models to only tens per year in the strong feedback models at a redshift of ten. The event rate therefore provides a potentially powerful observable signature of the physical processes regulating early BH growth. However, we must be cautious as degeneracies due to the seeding process will act here too.
    
    \item LISA and LGWA provide complementary windows into the population. LISA is the most sensitive to the $10^5-10^8$ \msolar chirp mass systems in our simulations while LGWA probes a lower-mass portion of the population. A subset of mergers is potentially detectable by both observatories, demonstrating the possibility of complementary GW observations of early BH assembly.

\end{itemize}

Overall, our results demonstrate that BH feedback does more than regulate gas accretion: it regulates the hierarchical assembly pathway of BHs by controlling whether newly formed seeds can grow sufficiently to become active participants in repeated mergers. Coupled with uncertainties in our seeding choices and other BH subgrid prescriptions this leaves us with multiple degeneracies that will need to be broken in advance of LISA and other next generation GW facilities. Current and future observations with JWST and other high-z electromagnetic instruments (ideally including X-ray) will be required to help break degeneracies.

\section*{Acknowledgements}
\noindent JR acknowledges support from the Royal Society and Research Ireland under grant number 
 URF\textbackslash R1\textbackslash 191132. LP, DM \& JR acknowledge support from the Research Ireland Laureate programme under grant number IRCLA/2022/1165. JHW acknowledges support from NSF grants AST-2108020 and AST-2510197 and NASA grant 80NSSC21K1053. RSB acknowledges support from a UKRI Future Leaders Fellowship MR/Y015517/1.
 \ \
The simulations were performed on the Luxembourg national supercomputer MeluXina and the Czech Republic EuroHPC machine Karolina.
The authors gratefully acknowledge the LuxProvide teams for their expert support.
\ \ 
The authors wish to acknowledge the Irish Centre for High-End Computing (ICHEC) for the provision of computational facilities and support.
\ \
The authors acknowledge the EuroHPC Joint Undertaking for awarding this project access to the EuroHPC supercomputer Karolina, hosted by IT4Innovations through a EuroHPC Regular Access call (EHPC-REG-2023R03-103) and to the LuxProvide supercomputer Meluxina  through a EuroHPC Regular Access call (EHPC-REG-2025R01-008).
\ \
SK has been supported by a Research Fellowship from the Royal Commission for the Exhibition of 1851.
\ \ 
MAB is supported by a UKRI Stephen Hawking Fellowship (EP/X04257X/1).
\ \ 
DS acknowledges support from the Science and Technology Facilities Council (STFC) under grant ST/W000997/1.
\ \
RSK and SCOG acknowledge financial support from the European Research Council via the ERC Synergy Grant ``ECOGAL'' (project ID 855130),  from the German Excellence Strategy via the Heidelberg Cluster of Excellence (EXC 2181 - 390900948) ``STRUCTURES'', from the German Science Foundation under grant KL 1358/22-1, and from the German Federal Ministry for Economic Affairs and Energy in project ``MAINN'' (funding ID 50OO2206). RSK is grateful for computing resources provided by the Ministry of Science, Research and the Arts (MWK) of the State of Baden-W\"{u}rttemberg through bwHPC and the German Science Foundation (DFG) through grants INST 35/1134-1 FUGG and 35/1597-1 FUGG, and also for data storage at SDS@hd funded through grants INST 35/1314-1 FUGG and INST 35/1503-1 FUGG. RSK also thanks for computing time provided by the Leibniz Rechenzentrum via grants pr32lo, pr73fi and GCS large-scale project 10391. 
\ \ 
PCC acknowledges support from a STFC Small Award (UKRI1187): ``Probing the origins of stars and life with a new approach to chemical modelling''.
\ \
MGH has been supported by STFC consolidated grants ST/N000927/1 and ST/S000623/1.

\bibliographystyle{aasjournal}
\bibliography{SEEDZ_References}



\end{document}